\documentclass[aps,prb,twocolumn,superscriptaddress,floatfix,longbibliography]{revtex4-2}

\usepackage{amsmath,amssymb} 
\usepackage{physics}
\usepackage{bm} 
\usepackage [autostyle, english = american]{csquotes}
\MakeOuterQuote{"}

\usepackage{enumitem}
\usepackage{graphicx} 
\usepackage{comment} 
\usepackage{textcomp} 
\usepackage{enumitem}
\setlist{noitemsep,leftmargin=*,topsep=0pt,parsep=0pt}

\usepackage{xcolor} 
\definecolor{lightgray}{gray}{0.6}
\definecolor{medgray}{gray}{0.4}

\usepackage{braket}
\newcommand{\sigmaz}{\hat{\sigma}^z}
\newcommand{\sigmay}{\hat{\sigma}^y}
\newcommand{\sigmax}{\hat{\sigma}^x}

\newcommand{\tz}{\hat{\tau}_3}

\newcommand{\tx}{\hat{\tau}_1}

\newcommand{\Hhat}{\hat{H}}

\newcommand{\U}{\hat{U}}
\newcommand{\+}{\dagger}

\newcommand{\h}{\hbar}

\newcommand{\kp}[1]{\ket{\psi #1 }}
\newcommand{\mbf}[1]{\mathbf{#1}}

\usepackage{hyperref}
\hypersetup{
colorlinks=true,
urlcolor= blue,
citecolor=blue,
linkcolor= blue,
}

\newcommand{\mytitle}{Suppressing excitations using quantum-Brachistochrone and nearest-neighbour interactions}

\begin{document}

\title{\mytitle}
\author{S John Sharon Sandeep}
\email[]{sjohn20@iiserb.ac.in}
\affiliation{Department of Physics, IISER Bhopal}

\author{Dibyajyoti Sahu}
\email[]{dibyajyoti20@iiserb.ac.in}
\affiliation{Department of Physics, IISER Bhopal}

\author{Suhas Gangadharaiah}
\email[]{suhasg@iiserb.ac.in}
\affiliation{Department of Physics, IISER Bhopal}
\date{\today}

\begin{abstract}
    We examine excitation suppression in the transverse-field Ising model (TFIM), where finite-time drive across a quantum critical point is assisted by the presence of a time-dependent coupling parameter. While conventional counterdiabatic protocols are designed to eliminate excitations, they often require complex many-body terms that are difficult to realize experimentally. 
    In contrast, our approach employs a local, time-dependent modulation of an existing coupling term in the Hamiltonian. 
    Within the framework of quantum optimal control, we find that under a linear ramp of the transverse field, the optimal evolution of the second parameter follows a non-monotonic trajectory. 
    For the TFIM, this protocol yields higher fidelity and improved robustness against noise compared to several orders of approximate counterdiabatic driving. 
    Furthermore, we provide an analytical demonstration of anti-Kibble–Zurek scaling in the presence of noise acting on either  the transverse field or the longitudinal coupling. 
    These results highlight the potential of this approach for developing simple, noise-resilient protocols for finite-time quantum state preparation.
\end{abstract}
\maketitle
\section{\label{sec:Introduction}Introduction}
Adiabatic driving is essential for the realization of quantum technologies such as 
quantum simulation and adiabatic quantum 
computation (AQC)~\cite{AQC, perspectivesofQA}. Effective operation relies on maintaining 
the system within the ground state manifold throughout the process. 
However, this approach encounters an unavoidable barrier due to an 
exponentially small energy gap between the ground state and excited 
states, a common feature near quantum phase transitions (QPTs). This 
vanishingly small gap necessitates extremely slow driving speeds to avoid 
excitation into higher states, leading directly to an exponential increase 
in computation 
times~\cite{QAshowsexponentially, kato1950adiabatic,kibble1976topology, zurek1985cosmological,dutta2010transverse, mukherjee2007quenching, zurek2005dynamics, cherng2006entropy, dziarmaga2005dynamics, polkovnikov2005universal, sen2008defect}.

A major challenge with slow drive or longer annealing times is the limited system decoherence time and the susceptibility to noise inherent in the system parameters. This constraint requires that the entire experiment be completed before the quantum information is irreversibly degraded, thereby forcing the system away from the adiabatic regime~\cite{decoherenceinQA}. To circumvent this difficulty, several protocols were proposed to exploit or engineer a spectral gap that allows for faster driving of the Hamiltonian while still preserving adiabaticity. Some of these include diabatic quantum annealing \cite{crosson2021prospects}, optimal nonlinear passage across a QCP \cite{PhysRevLett.101.076801, sen2008defect}, optimal quantum control strategies \cite{DAlessandro2021,Glaser2015}, etc.
Within this set of methods, counter-diabatic (CD) driving~\cite{berry_2009, demirplak2003adiabatic, delcampoSTAbyCD} has emerged as a systematic approach towards this goal and falls under the broader umbrella of shortcut to adiabaticity~\cite{delcampoSTAbyCD,guery2019shortcuts,ExperimentalLCD}.
In CD driving, the system Hamiltonian is supplemented by an auxiliary term designed to ensure that the evolving state exactly follows the instantaneous ground state of the original Hamiltonian. While this framework has been theoretically established across a wide range of models~\cite{CDmodels1, CDmodels2, CDmodels3}, its experimental implementation remains difficult. The main challenge lies in the fact that the auxiliary term often involves complicated operator structures, including nonlocal and multispin interactions~\cite{delcampo2014IsingModel, variational2}. In addition to this structural complexity, the auxiliary term can only be derived by requiring complete information about the instantaneous eigenstates of the system Hamiltonian at all times~\cite{2parameterCD,cold2023counterdiabatic}. To address these challenges, Ref.\cite{polkovnikov2017minimizing, polkovnikov2017geometry} developed a variational approach that systematically constructs approximate CD Hamiltonians. This strategy identifies local and experimentally accessible forms of driving, which have since been demonstrated in both theoretical models\cite{LCDmodels1, LCDmodels2, LCDmodels3} and experimental platforms~\cite{ExperimentalLCD}. An alternative route is truncated CD driving~\cite{delcampo2014IsingModel, damski2014counterdiabatic}, where the exact CD Hamiltonian is systematically expanded in terms of long-range interactions and only terms up to a chosen order are retained. This controlled truncation suppresses a significant fraction of excitations while avoiding the full complexity of nonlocal multispin terms, thereby striking a balance between accuracy and experimental feasibility. An improvement over this approach was realized by the introduction of an additional time-dependent driving function in the Hamiltonian with a corresponding two-parameter
CD Hamiltonian \cite{2parameterCD} which was variationally derived via a method similar to the case of a single parameter CD. This method was shown to enhance the final ground-state fidelity. It was also shown analytically that for a general class of integrable systems having a second parameter in the Hamiltonian can be advantageous in suppressing excitations in the system~\cite{sau2014suppressing}.\

In prior works on two-parameter driving, the time dependence of the second parameter was typically chosen in an \textit{adhoc} manner.
An alternate strategy for further excitation suppression is the Counterdiabatic Optimized Local Driving (COLD) approach \cite{cold2023counterdiabatic, Barone_2024}, which combines local CD driving with optimal control. While many of the earlier studies employed fidelity-based cost functions, Čepaitė et al.~\cite{cold2023counterdiabatic} proposed minimizing the variance of the CD Hamiltonian as an alternative optimization strategy. Building on this idea, we employ the variance of the CD Hamiltonian as a cost functional to optimize the additional control fields. As we demonstrate here, this strategy also extends the results of Ref.~\cite{sau2014suppressing} to cases with finite drive times and boundary conditions imposed at finite times. Previous studies \cite{delcampo2014IsingModel, polkovnikov2017minimizing} have shown that while the local approximate CD driving can improve fidelity; such protocols often fail to deliver computationally useful fidelities for an extended range of drive rate, particularly in the presence of noise. To illustrate this limitation, we consider a $50$-site transverse-field Ising model (TFIM) driven across a single critical point, where CD corrections even up to the tenth order fail to achieve  high ground-state fidelity.  In contrast, we demonstrate that introducing a second control parameter in the longitudinal coupling terms with an optimized time profile can substantially enhance the performance. Specifically, this two-parameter protocol tolerates control noise-strength up to $\sim 1/N$ while still achieving a high fidelity of $\sim 0.96$, suggesting that this approach merits consideration.\

Intuitively, such optimized protocols operate by identifying the adiabatic path characterized by a large energy gap and a reduced drive rate near the critical point  \cite{cold2023counterdiabatic}. The enhanced gaps at the criticality offer intrinsic robustness to noise, up to a limit set by the minimum energy gap~\cite{jordan2006error, bookatz2015error}. We make this intuition concrete  by analyzing the Hamiltonian with the added control term in the momentum space. We explicitly show that for modes near the critical point, the energy gap increases while the effective drive rate decreases. 
More generally, approximating the additional time-dependent functional by a quadratic form quadratic allows us to derive compact analytical expressions for the effective drive rate and the minimum gap. 
These results yield analytical estimates of the excitation density for both the noisy and noiseless scenarios, clarifying the role of optimized control in suppressing non-adiabatic excitations.
In all the simulations presented in this work, we consider dynamics in the presence of noise acting on a single control parameter at a time, either in the transverse field or the longitudinal coupling. 
In both the cases, we find that noise induces an Anti-Kibble–Zurek-type scaling ($n \propto 1/v$).
Finally, by combining the noisy results with our analysis of excitation density in the noiseless case, we demonstrate that the optimal drive rate can be analytically estimated in the presence of noise in either control field.

The remainder of the paper is organized as follows. Section~\ref{sec:model} introduces the models central to our study, with a focus on the transverse-field Ising chain and its variants. In Sec.~\ref{sec:CD}, we revisit the CD driving approach, highlight its limitations, and motivate the need for an alternative methodology. In section~\ref{sec:optimization} we develop a general optimization strategy based on the quantum brachistochrone formalism. Analytical results for the optimal protocol and a scaling law for the diabatic error are presented in Sec.~\ref{sec:Analytical_scaling}. In Sec.~\ref{sec:Noise}, we examine the effect of classical noise on the performance of the optimized protocol, followed by the derivation of excitation density under noisy conditions in Sec.~\ref{subsec:Analytical_scaling_noise}. Finally, Sec.~\ref{sec:conclusion} summarizes our main findings and outlines future directions.

\section{\label{sec:model}Model}
A particularly important class of models in the study of non-equilibrium quantum dynamics is that of translationally invariant free-fermion systems. Owing to their analytical tractability, such models admit exact solutions thus helping to reveal the essential features of non-adiabatic transitions and defect formation during time-dependent evolution.
Keeping this in mind, we consider the following free-fermion Hamiltonian:
\begin{equation}\label{Models eq: H}
    H(t) = \sum_k \psi_k^\dagger H_k(t) \psi_k,
\end{equation}
where $\psi_k^\dagger = (c_{1k}^\dagger, c_{2k}^\dagger)$ are Fermionic creation operators, and $H_k(t)$ is given by~\cite{sau2014suppressing}
\begin{equation}\label{Models eq: model}
    H_k(t) = \tau_3 \left[\lambda_1(t) - \lambda_2(t)b_k\right] + \tau_1 \lambda_3(t) g_k ,
\end{equation}
Here, $\tau_3$ and $\tau_1$ are Pauli matrices, while $b_k$ and $g_k$ are model-dependent functions of momentum $k$. The parameters $\lambda_1$,$\lambda_2$, and $\lambda_3$ are time-dependent drive parameters whose specific forms depend on the model and driving protocol.

Earlier studies have analyzed the dynamics of this Hamiltonian under the quench of a single parameter~\cite{mukherjee2007quenching, zurek2005dynamics,cherng2006entropy,dziarmaga2005dynamics,polkovnikov2005universal, sen2008defect}, resulting in the standard Kibble Zurek scaling with one control parameter. Reference~\cite{sau2014suppressing} extended this analysis to simultaneous quenches of $\lambda_1$ and $\lambda_3$, demonstrating improved scaling of the ground state probability with the drive rate.
In this  work, we consider finite-time drive protocol a scenario more relevant for quantum annealing applications.
Within this framework, we apply the techniques introduced in Ref.~\cite{cold2023counterdiabatic} to determine the optimal time dependence of the parameters $\lambda_i(t)$.

To make the discussion concrete, we now focus on the one-dimensional TFIM, driven through its quantum critical point at a finite rate. This model is also chosen due to its relevance in quantum annealing applications. The Hamiltonian is given by
\begin{equation}\label{eq: Ho}
    \hat{H}^0(t) =  -\hbar J \sum_i \left( g(t)\, \sigma^z_i + \sigma^x_i \sigma^x_{i+1} \right),
\end{equation}
where $g(t)$ is the time-dependent transverse field. We choose units such that $\hbar = 1$ and $J = 1$. The system is driven through the quantum critical point at $g_c = 1$, by varying $g(t)$ from an initial value $g_i \gg 1$ at time $t = 0$ to a final value $g_f = 0$ at a finite time $T$. We define the drive velocity as \( v = (g_i - g_f)/T \), which sets the drive rate. If this variation is slower than a characteristic rate determined by the energy gap at the critical point—starting from the ground state of $\hat{H}^0(0)$ it is possible to reach the ground state of the final Hamiltonian $\hat{H}^0(T)$ with high probability.

Since the global $\mathbb{Z}_2$ parity operator $\hat{P} = \prod_i \sigma^z_i$ is conserved, and the system is initialized in the ground state (which lies in the $+1$ parity sector), we restrict our analysis to this subspace.
Applying the projection and performing a Jordan-Wigner transformation~\cite{jordan1928paulische,lieb1961two}, the spin operators are mapped to spinless fermions:
\begin{equation}
    \sigma^z_i = 1 - 2 c_i^\dagger c_i, \quad \sigma^+_i = c_i \prod_{j < i} (1 - 2 c_j^\dagger c_j),
\end{equation}
acquiring the following form
\begin{equation}
    \hat{H}^0(t) = \sum_i \left[ -2 g(t)\, c_i^\dagger c_i + \left( c_i^\dagger c_{i+1} + c_i^\dagger c_{i+1}^\dagger + \text{h.c.} \right) \right],
\end{equation}
with antiperiodic boundary conditions on the fermions.

We now perform a Fourier transform consistent with these antiperiodic conditions:
\begin{equation}
    \hat{c}_j = \frac{e^{-i\pi/4}}{\sqrt{N}} \sum_k \hat{c}_k e^{ikj},
\end{equation}
where $k \in \left\{ \pm \frac{(2n - 1)\pi}{N} ~\big|~ n = 1, 2, \dots, N/2 \right\}$. This transformation leads to a block-diagonal form of the Hamiltonian, Eq.~\eqref{Models eq: H}, with
\begin{equation}\label{Models eq: model TFIM}
    H_k^0(t) = 2 \left[ g(t) - \cos(k) \right] \tau_3 + 2 \sin(k)\, \tau_1,
\end{equation}
which matches with the form introduced earlier in Eq.~\eqref{Models eq: model} with the identifications $\lambda_1(t) = 2g(t)$, $\lambda_2(t) = 2$, $\lambda_3(t) = 2$, $b_k = \cos(k)$, , and $g_k = \sin(k)$.

The dynamics are quantified via two observables, the defect density and the diabatic error. Defect density is a measure of the the average number of excitations generated during the evolution. It is defined as
\begin{equation}
    n_{ex} = \frac{1}{N} \sum_{k>0} p_k,
\end{equation}
where $p_k$ is the probability that the $k$th mode is excited at the end of the drive. The second quantity is the \textit{diabatic error}, which captures the total probability of not being in the many-body ground state at the end of the protocol. It is given by
\begin{equation}
    \text{DE} = 1 - \mathcal{F} = 1 - \prod_{k>0} \left(1 - p_k\right),
\end{equation}
where $\mathcal{F} = \prod_{k>0} (1 - p_k)$ is the probability that all modes remain in their respective instantaneous ground states. These two observables serve as complementary diagnostics of non-adiabatic effects during the time evolution.\
\section{\label{sec:CD}Suppressing Non-adiabatic excitations}
When the Hamiltonian is driven through one of its critical points, namely at $ g = \pm 1 $, non-adiabatic transitions become significant if the quench rate exceeds the minimal energy gap. This leads to the generation of quasiparticle excitations, thereby reducing the fidelity of the evolved state. A systematic method to suppress such excitations is CD driving, where the Hamiltonian is modified by an auxiliary term designed to guide the system along its instantaneous eigenstates. For the transverse field Ising chain, the exact CD term in real space is given by
\begin{equation}\label{eq: CD real space Hamiltonian}
    \hat{H}^0_{\mathrm{CD}} = \dot{g} \sum_{m=1}^{\frac{N}{2}-1} h_m(g)\, \mathcal{C}^{[m]} + \frac{1}{2} h_{N/2}(g)\, \mathcal{C}^{[N/2]},
\end{equation}
where
\[
    \mathcal{C}^{[m]} = \sum_{n=1}^{N} \left[ \sigma_n^x \sigma_{n+m}^y + \sigma_n^y \sigma_{n+m}^x \right] \left( \prod_{j=n+1}^{n+m-1} \sigma_j^z \right),
\]
and $ h_m(g) $ are functions of the transverse field $ g $, with explicit expressions provided in Appendix~\ref{app:CD} and Refs.~\cite{delcampo2014IsingModel, damski2014counterdiabatic}.

\begin{figure}[th]
    \includegraphics[width=\linewidth]{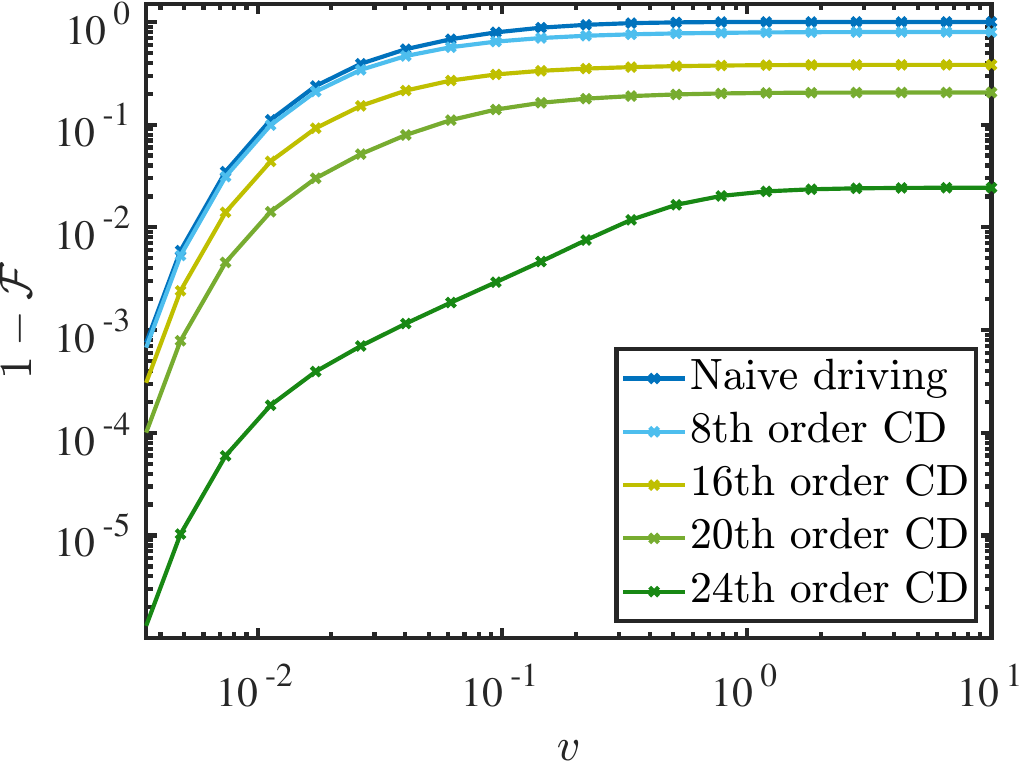} 
    \caption{Excitation probability $1-\mathcal{F}$ versus quench rate $v$ for the naive and the truncated counterdiabatic driving (with cutoffs $M=8,16,20,24$) protocol. Increasing $M$ drives the evolution closer to the adiabatic limit, while the fidelity saturates at large $v$. Parameters: $N=50$, $g_i=10$, $g_f=0$.}
    \label{fig: CD 1600 sites}
\end{figure}

The CD Hamiltonian consists of long-range multispin interactions extending up to $ N/2 $ sites (e.g., 25 sites for $ N = 50 $), making it difficult for experimental implementation~\cite{damski2014counterdiabatic}. A practical approach involves employing a \emph{truncated} counterdiabatic Hamiltonian, in which interactions are considered only up to a finite range $M$. Using the exact form of $ h_m(g) $ for finite system size $ N $, we numerically evaluate the performance of the truncated CD Hamiltonian for various values of $M$. As shown in Fig.~\ref{fig: CD 1600 sites} even a 16th-order truncated CD Hamiltonian does not achieve fidelities higher than 0.9.

The limited improvement in fidelity achieved by truncated CD driving in system with finite-size motivates the exploration of alternative strategies. Rather than relying on long-range CD corrections, we consider augmenting the Hamiltonian with control terms composed of operators already present in the real-space formulation. This approach is motivated by two key considerations: first, such operators correspond to local, experimentally accessible interactions; and second, their inclusion preserves the analytical tractability of the model.

Specifically, we modify the annealing Hamiltonian by introducing a time-dependent term of the form:
\begin{equation*}
    f(t) \sum_i \sigma_i^x \sigma_{i+1}^x,
\end{equation*}
where the auxiliary control function \( f(t) \) satisfies the boundary conditions \( f(0) = f(T) = 0 \). This ensures that the system evolves between the same initial and final Hamiltonian while allowing for a more flexible trajectory in parameter space.
This design choice closely resembles reverse or diabatic annealing approaches~\cite{crosson2021prospects}, in which an auxiliary Hamiltonian is introduced such that it vanishes at the boundaries of the protocol but plays an active role during intermediate times. For this reason, we use the terms “optimal driving” and “diabatic driving” interchangeably throughout this work to describe our two-parameter strategy. Although one could, in principle, include the other local operators such as \( \sigma_i^y \sigma_{i+1}^y \), we restrict our attention to the \( \sigma_i^x \sigma_{i+1}^x \) term. This choice is motivated by the fact that the latter term is is already present in the original Hamiltonian, making it experimentally accessible. In this case, the total Hamiltonian can again be brought into the form of Eq.~\eqref{Models eq: H}, with \( H_k(t) \) given by
\begin{align}
    \Hhat'_k(t) & = \tz \left[2 g(t) - 2\left(f(t) + 1\right) \cos(k) \right] \nonumber \\
                & \hspace{5em} + \tx\, 2\left(f(t) + 1\right) \sin(k),
    \label{eq: H_k}
\end{align}
\noindent
which retains the structure of decoupled two-level systems. We will show that, when optimized appropriately, this two-parameter protocol can achieve significantly lower excitation than naive linear driving and can even surpass standard truncated CD protocols.

\section{\label{sec:optimization}Optimization}
To optimize the auxiliary control field \( f(t) \), we draw upon variational principle strategies (Refs.~\cite{cold2023counterdiabatic, suzuki2020performance}) for designing time-dependent
protocols. Before introducing the general optimization procedure, we first explore the commonly used quadratic form~\cite{crosson2021prospects}
\begin{equation}\label{eq: alpha diabatic pulse}
    f(t)=\alpha \frac{t}{T} (1 - \frac{t}{T}) ~~~\text{where,}~\alpha > 0.
\end{equation}
While it is not the most optimal choice, the appeal of this form lies in its simplicity.
To guide our optimization, we focus on drive rates for which only the low-momentum modes are excited. In this regime, the mode-resolved Hamiltonian can be approximated as
\begin{equation}\label{eq: XY fermionic Hamil, k near 0}
    \hat{H}'_k(t) \approx 2 \,
    \begin{pmatrix}
        g(t) - f(t) - 1 & (f(t) + 1)\, k   \\
        (f(t) + 1)\, k  & -g(t) + f(t) + 1
    \end{pmatrix}.
\end{equation}
The critical time \( t_c \), can be approximated as the time for which the coefficient of \( \tau_3 \) in Eq.~\eqref{eq: XY fermionic Hamil, k near 0} vanishes. The main effect of introducing the slow varying control term \( f(t) \) is to shift the critical time. In the naive case (i.e., $f(t) = 0$), this occurs when \( g(t_c) = 1 \), but with the control term, it satisfies \( g(t_c) = f(t_c) + 1 \). Moreover, the presence of \( f(t) \) also increases the minimum instantaneous gap from \( 2k \), in the naive case, to \( 2k(f(t_c) + 1) \). Thus, the effectiveness of this control protocol is primarily governed by the value of \( f(t) \) near \( t_c \). Importantly, the choice of \( \alpha \) in Eq.~\eqref{eq: alpha diabatic pulse} cannot be made arbitrarily large. For fixed values of \( g_i \) and \( g_f \), increasing \( \alpha \) leads to a steeper \( f(t) \), which enhances \(  \dot{H}(t) \) and consequently increases non-adiabatic transitions. A quick approach is to select \( \alpha \) such that \( t_c \) aligns with the midpoint of the protocol, i.e., \( t_c = T/2 \). This allows sufficient time for \( f(t) + 1 \) to reach the target value \( g(T/2) \). Incidentally this choice also coincides with the maximum value of $f(t)$ and thus the slowest rate of change of control term. For \( f(t) = \alpha \frac{t}{T} (1 - \frac{t}{T}) \), solving \( g(T/2) - f(T/2) - 1 = 0 \) yields
\(
\alpha = 4(g(T/2) - 1)  = 16
\)
for a linear ramping of $g$ from \( g_i = 10 \) to \( g_f = 0 \). The resulting control function is shown (in yellow) in Fig.~\ref{fig: pulses_fullquench}.
In the following, we construct a sophisticated control function that outperforms this simple quadratic pulse.

\subsection{\label{subsec:QBC}Solution via Quantum Brachistochrone}

In the above analysis, two main considerations guided our search for an effective control function \( f(t) \): (i) a slow rate of change of the Hamiltonian, and (ii) an increased energy gap at the critical point. The following expression captures both of these aspects succinctly:
\begin{equation}\label{eq: Lagrangian 1}
    \mathcal{L}[f(t),\dot{f}(t)] = \|\dot{f}(t) \frac{\partial \Hhat}{\partial f}(t)\|/\Delta^2(t)
\end{equation}
where \( \Delta(t) \) denotes the energy gap between the ground state and the first excited state of \( \hat{H}(t) \) and the norm is the Hilbert-Schmidt norm defined as $\|A\| = \sqrt{Tr[A^\dagger A]}$. This quantity was first introduced in Ref.~\cite{rezakhani2009quantum} as the adiabatic time functional. It not only quantifies adiabaticity but also enables optimization of the time-dependent control \( f(t) \) by minimizing the time-integral of \( \mathcal{L}[f(t),\dot{f}(t)] \).

In this work, we adopt a different cost functional for quantifying adiabaticity: the variance of the CD Hamiltonian associated with \( \hat{H}(t) \). This approach was introduced in the context of quantum speed limits (QSL) in Ref.~\cite{suzuki2020performance}, where it was used to establish a lower bound on the fidelity. Using the Mandelstam-Tamm (MT) relation \cite{mandelstam1945uncertainty}:
\begin{equation}\label{eq: MT relation}
    \cos^{-1} \left( \braket{\psi(0)|\psi(T)} \right) \leq \int_0^T dt\, \sigma(\hat{H}(t), \ket{\psi(t)}),
\end{equation}
where \( \sigma(\hat{O}, \ket{\psi}) = \sqrt{\bra{\psi}\hat{O}^2\ket{\psi} - (\bra{\psi}\hat{O}\ket{\psi})^2} \) and \( \ket{\psi(t)} \) is the time-evolved state at time \( t \), an alternate bound was derived in Ref.~\cite{suzuki2020performance} given by
\begin{equation}\label{eq: MT Fidelity}
    \begin{aligned}
        \cos^{-1} \left( \braket{\psi_{\mathrm{ad}}(T)|\psi(T)} \right) \leq \min \Bigg\{
         & \int_0^T dt\, \sigma(\hat{H}_{\mathrm{CD}}(t), \ket{\psi(t)}),              \\
         & \int_0^T dt\, \sigma(\hat{H}_{\mathrm{CD}}(t), \ket{\psi_{\mathrm{ad}}(t)})
        \Bigg\}
    \end{aligned}
\end{equation}
Here, \( \ket{\psi_{\mathrm{ad}}(t)} \) denotes the instantaneous ground state of \( \hat{H}(t) \), and \( \hat{H}_{\mathrm{CD}}(t) \) is the corresponding CD Hamiltonian for \( \hat{H}(t) \). From Eq.~\eqref{eq: MT Fidelity}, it follows that reducing the instantaneous variance of \( \hat{H}_{\mathrm{CD}}(t) \) improves the lower bound of the fidelity. This naturally links the problem of designing optimized control protocols to the broader framework of the Quantum Brachistochrone (QB), where one seeks the shortest possible path consistent with physical constraints. Motivated by this perspective, we numerically optimize the control function \( f(t) \) by minimizing the time-integrated variance:
\[
    \int_0^T dt\, \sigma(\hat{H}_{\mathrm{CD}}'(t), \ket{\psi_{\mathrm{ad}}(t)}).
\]
Since the Hamiltonian in Eq.~\eqref{Models eq: H} is exactly diagonalizable, the variance can be written in the form of the Hilbert-Schmidt norm (see Appendix \ref{app:Hilbert}):
\[
    \sigma(\hat{H}_{\mathrm{CD}}'(t), \ket{\psi_{\mathrm{ad}}(t)}) = \| \hat{H}'_{\mathrm{CD}}(t) \|/\sqrt{2}.
\]
While we use the variance as the cost functional, the Hilbert-Schmidt norm can be used as a practical substitute when evaluating the full variance is unfeasible~\cite{cold2023counterdiabatic}. The choice of evaluating variance in the adiabatic basis, rather than in the evolved basis, is motivated by practical considerations. In this basis, the variance can be computed without explicitly evolving the Hamiltonian, enabling significantly faster optimization while providing a broadly applicable cost functional that can serve as an alternative to fidelity-based cost functional.

\begin{figure}[t]
    \centering
    \includegraphics[width=\linewidth]{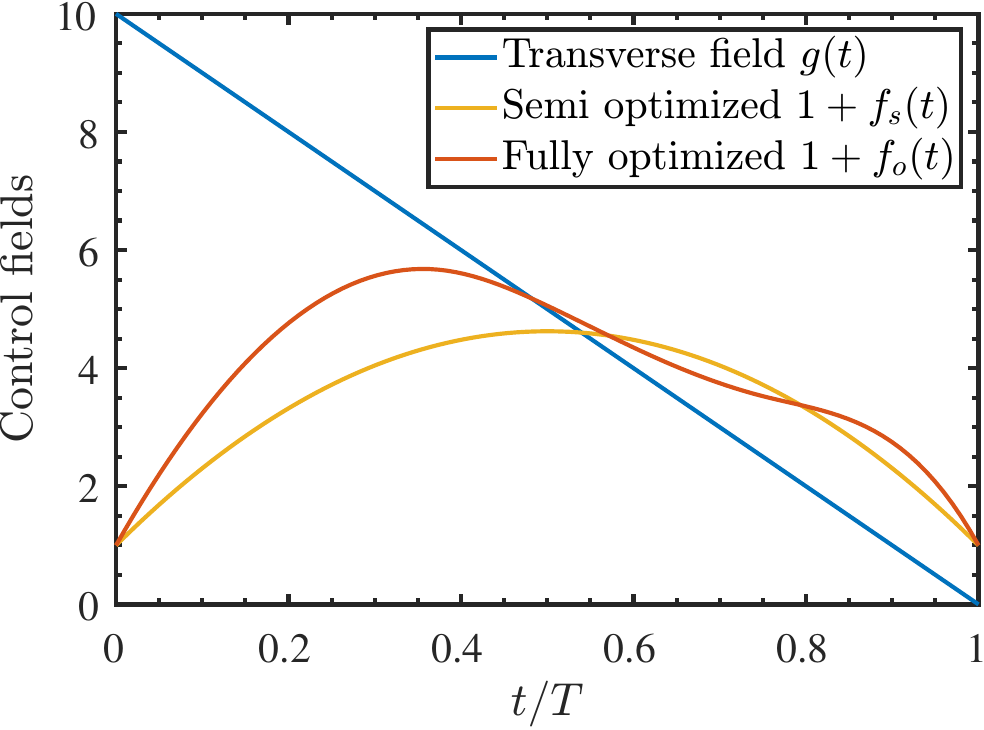}
    \caption{Transverse field $g(t)$ and control pulse $1+f(t)$ for the semi-optimized and fully optimized protocols.}
    \label{fig: pulses_fullquench}
\end{figure}

Minimizing the time-integrated cost yields the optimized control function \( f_o(t) \), shown in Fig.~\ref{fig: pulses_fullquench}. We refer to this as the \textit{fully optimized} protocol, since the entire time profile of \( f(t) \) is obtained variationally without assuming a fixed functional form. For comparison, we also consider a simpler ansatz of the form \( f_s(t) = \alpha \frac{t}{T}(1 - \frac{t}{T}) \), which we refer to as the \textit{semi optimized} protocol. Here, only the parameter \( \alpha \) is optimized while keeping the functional form fixed. Both protocols satisfy the same boundary conditions but differ notably in their time dependence, particularly near the critical point where the counterdiabatic contribution is most significant. For the linear ramp of \( g(t) \) considered in this work, a variational solution obtained via Python’s \texttt{SciPy} package gives \( \alpha \approx 14.1 \), closely matching the earlier heuristic estimate of \( 16 \). It is to be noted that it is enough to solve for $f(t)$ for a single drive-rate $v=1$, whose solution we shall denote by $f_{o/s}(t)$, so that the solution for a general drive rate $v$ is simply given by $f_{o/s}(vt)$.

\begin{figure}[t]
    \centering
    \centering
    \includegraphics[width=\linewidth]{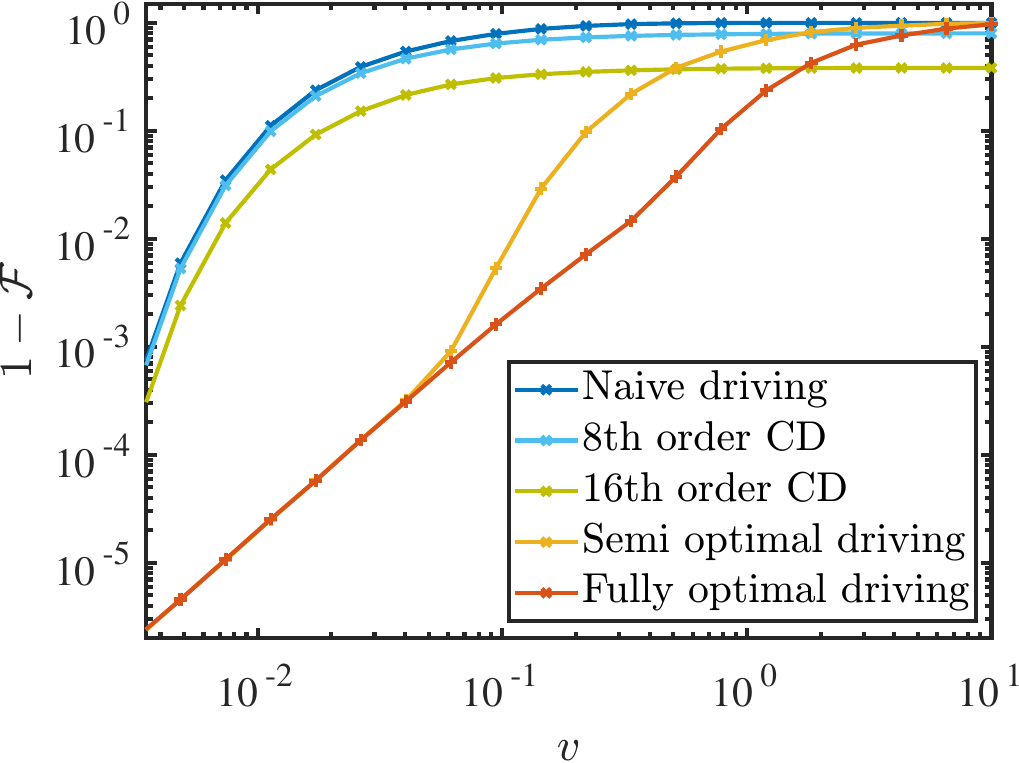}
    \caption{Excitation probability vs.\ drive rate under a linear ramp of the transverse field from $g=10$ (paramagnetic) to $g=0$ (ferromagnetic), via different protocols. Both semi-optimized and fully optimized protocols acquire significantly lower excitations as compared to the naive and truncated CD protocols over a large range of drive rate.}
    \label{fig: fVsv_linear}
\end{figure}

\begin{figure}[t]
    \centering
    \centering
    \includegraphics[width=\linewidth]{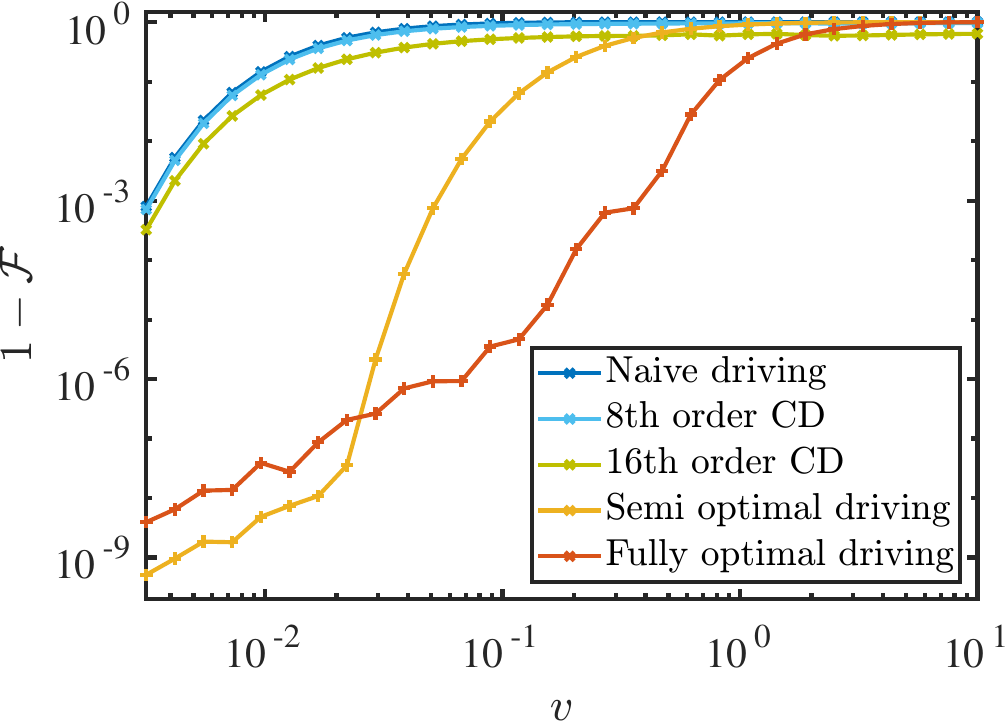}
    \caption{Excitation probability vs.\ drive rate under a linear ramp of the transverse field from $g=10$ to $g=-10$ (paramagnetic to paramagnetic), assisted by different protocols. Once again, semi-optimized and fully optimized protocols acquire significantly lower excitations as compared to the naive and truncated CD protocols over a large range of drive rate.}
    \label{fig:fVsv_linear_full}
\end{figure}

We next examine the fidelity achieved under the optimized annealing scheme and compare it with that obtained using approximate CD driving of various orders. As shown in Fig.~\ref{fig: fVsv_linear}, both diabatic protocols, the \textit{semi-optimized} (yellow curve) and the \textit{fully optimized} (red curve), outperform multiple orders of approximate CD driving.
Besides the half-quench case, we also consider full quench with a linear ramp of \( g(t) \) from \( g_i = 10 \) to \( g_f = -10 \). For convenience, we redefine the time interval such that \( t \in [-T/2, T/2] \), where the total drive time \( T \) is related to the drive rate by \( |v| = (g_i - g_f)/T \).
We use the quadratic control function introduced earlier in Eq.~\eqref{eq: alpha diabatic pulse}, now modified to match the new time interval:
\begin{equation}
    f_s(t) = -\alpha \left( \frac{t}{T} \right)^2 + \frac{\alpha}{4}.\label{eq:f_s_full}
\end{equation}
The optimal value of the parameter is found to be \( \alpha \approx 21.66 \). In contrast, the fully optimized path in control space deviates significantly from the quadratic form and instead exhibits an inverted double-well shape~\cite{numerical_optimi}. The corresponding control functions are shown in Fig.~\ref{fig: pulses_fullquench2}. As in the half-quench case, both the semi- and fully optimized protocols yield higher fidelities than multiple orders of truncated CD, as shown in Fig.~\ref{fig:fVsv_linear_full}.
The improvement in fidelity for both protocols can be attributed to two key features: (i) a widening of the energy gap at the critical point \( t_c \), and (ii) a reduced effective drive rate near \( t_c \). These aspects will be analyzed in detail in the following section.


\section{Analytical Results and Scaling Law} \label{sec:Analytical_scaling}
To gain analytical insight into the excitation dynamics, we employ the Landau–Zener (LZ) formula \cite{landau1932theory,zener1933dissociation}
, which provides an exact solution for a two-level system subjected to a linear quench from \( t = -\infty \) to \( t = \infty \). 
For this analysis, we consider the full-quench case, where the transverse field \( g(t) \) is ramped linearly from a large positive initial value \( g_i \gg 1 \) to a large negative final value \( g_f \ll -1 \). This ensures that the system begins and ends far from the critical region, aligning well with the asymptotic assumptions underlying the LZ solution. By mapping each momentum mode \( k \) of our system to an effective LZ problem, we can approximate the excitation probability of the corresponding quasiparticle. In contrast to the half-quench case, the full-quench setup permits a straightforward and faithful application of the LZ approximation.\

\begin{figure}[t]
    \centering
    \centering
    \includegraphics[width=\linewidth]{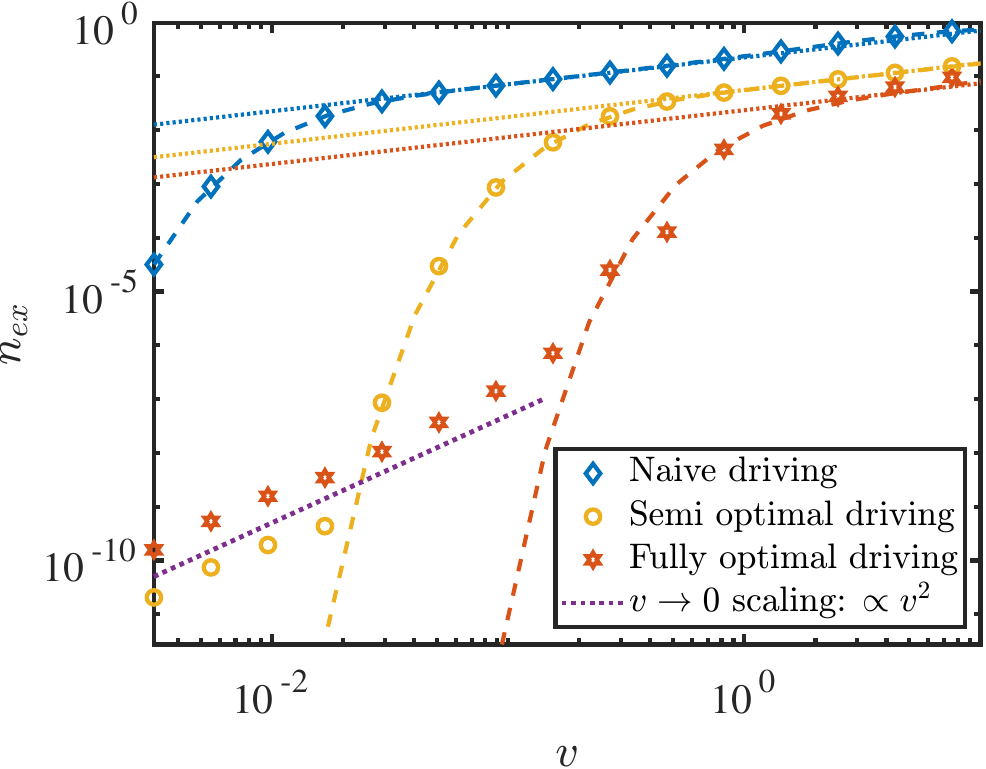}
    \caption{Density of excitations vs drive rate for various protocols under a linear ramp of the transverse field from $g_i = 10$ to $g_f = -10$. The points represent numerical simulation data. The dashed lines indicate the finite-size excitation density, while the dotted lines represent the thermodynamic-limit behavior in the slow to intermediate-driving regime.}

    \label{fig: densityVsv_linear_anal}
\end{figure}

We next briefly outline the derivation of the scaling of the excitation density
\(n_{\mathrm{ex}}\) with respect to the global drive rate \(v\).
The LZ formula gives $p = \exp ( -\pi \Delta^2/2 v)$ as the excitation probability for a two-level system with Hamiltonian \( H_{\text{LZ}}(t) = 1/2(\Delta \, \tau_1 + v t \, \tau_3) \), where \( \Delta \) is the minimum gap and \( v \) the sweep rate.
For naive protocol [\(f(t)=0\)] we identify from Eq.~\eqref{Models eq: model TFIM}, \( \Delta_k = 4 \sin k \) and \( v_k = 4v \), leading to the excitation probability
\[
    p_k = \exp\left( -\frac{2\pi \sin^2 k}{v} \right).
\]
For finite-size systems, the excitation density $n_{\text{ex}}$
can be directly computed by summing the excitation probability over momenta and is consistent with numerical data, as illustrated in Fig.~\ref{fig: densityVsv_linear_anal} (dashed line). In the thermodynamic limit and for intermediate to slow quenches, an analytical estimate of \( n_{\mathrm{ex}} \) can be obtained by integrating \( p_k \) over the momentum modes:
\[
    n_{\mathrm{ex}} \approx \frac{2}{\pi} \int_0^{\pi/2} \exp\left(-\frac{2\pi\,k^2}{v}\right) dk
    \;\approx\; \frac{\sqrt{v}}{\pi\sqrt{2}}.
\]
This expression recovers the well-known Kibble--Zurek scaling, where \( n_{\mathrm{ex}} \sim \sqrt{v} \). This scaling also agrees well with finite-size results in the intermediate drive-time regime (Fig.~\ref{fig: densityVsv_linear_anal} dotted line). For slow driving rates, however, contributions from near-zero-momentum modes are dominant leading to an exponential scaling.
We will show below that the Hamiltonian with optimized diabatic control preserves $\sqrt{v}$ scaling behavior, but with a notably smaller prefactor, indicating a suppressed excitation density.

\begin{figure}[t]
    \centering
    \includegraphics[width=\linewidth]{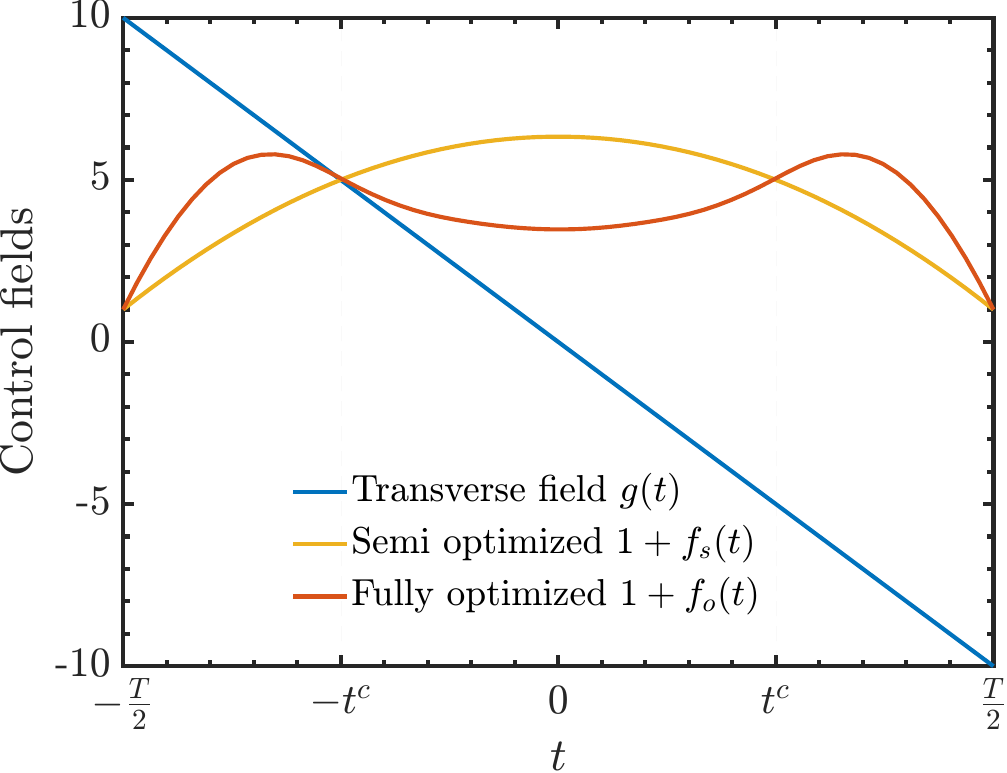}
    \caption{Transverse field $g(t)$ and control pulse $1+f(t)$ for the semi-optimized and fully optimized protocols under a full quench from $g_i=10$ to $g_f=-10$.}
    \label{fig: pulses_fullquench2}
\end{figure}

For the diabatic control pulse, we approximate the Hamiltonian for each \(k\)-mode as an effective two-level LZ problem, with a mode-dependent drive rate and energy gap~\cite{sau2014suppressing}. To identify the key parameters, we first determine the effective critical time \(t_k^c\), defined as the time for which the diagonal term of the Hamiltonian vanishes:
\begin{equation}\label{eq: LZT approx 1}
    g(t_k^c) - \left[ f(t_k^c) + 1 \right] \cos(k) = 0.
\end{equation}
The effective drive rate and the energy gap are obtained by linearizing the Hamiltonian around the critical time $t_k^c$ and are given by $v_k = 4 \left| g'(t_k^c) - f'(t_k^c) \cos(k) \right|$ and $\Delta_k = 4 \left[ f(t_k^c) + 1 \right] \sin(k)$, respectively.

For a general quadratic control field of the form \( \tilde{f}(t) = a(t/T)^2 + b \), the critical time, drive rate, and minimal energy gap for each \( k \)-mode are obtained analytically and are given by
\begin{equation}\label{eq: delta_k v_k fully-o pulse}
    \begin{aligned}
        \tilde{t}_k      & = T \frac{g_i \sec(k)}{a} \left( \sqrt{1 + \gamma \cos^2(k)} - 1 \right), \\
        \tilde{v}_k      & = 4v \sqrt{1 + \gamma \cos^2(k)},                                         \\
        \tilde{\Delta}_k & = \beta \sec^2(k) \left( \sqrt{1 + \gamma \cos^2(k)} - 1 \right) \sin(k),
    \end{aligned}
\end{equation}
where the parameters \(\gamma = -a(b + 1)/g_i^2\) and \(\beta = -8g_i^2/a\).
For the semi-optimized pulse [defined in Eq.~\eqref{eq:f_s_full}], we obtain
\[
    \gamma_s = \frac{\alpha^2 + 4\alpha}{4g_i^2}, \quad \beta_s = \frac{8g_i^2}{\alpha}.
\]

In contrast, the fully optimized pulse \(f_o(t)\) obtained numerically does not admit a closed-form expression. However, from Fig.~\ref{fig: pulses_fullquench2}, we observe that \(f_o(t)\) is approximately quadratic within the interval \((-t^c, t^c) \approx (-T/4, T/4)\), where \(\mp t^c\) are the critical times for the \(k \approx 0\) and \(k \approx \pi\) modes, respectively. Since the critical time \(t_k^c\) for each \(k\)-mode lies within this range, the excitation dynamics is effectively governed by the shape of \(f_o(t)\) in this interval. This motivates us to approximate \(f_o(t)\) by a quadratic function in this region.

To match the effective quadratic form \( \tilde{f}(t) \) with \( f_o(t) \), we determine the coefficients \(a\) and \(b\) by solving the conditions:
\[
    \left.\frac{d\tilde{f}}{dt}\right|_{-t^c} = \left.\frac{df_o}{dt}\right|_{-t^c}, \quad \tilde{f}(-t^c) = f_o(-t^c),
\]
The above constraints yield,
\[
    \gamma_o = \left(1 + \dot{f}_o^c\right)^2 - 1, \quad \beta_o = \frac{4(1 + f_o^c)}{\dot{f}_o^c},
\]
where \(\dot{f}_o^c = \left.\frac{df_o}{dt}\right|_{-t^c} / v\) and \(f_o^c = f_o(-t^c)\).

With the effective drive rate \(v_k\) and minimal energy gap \(\Delta_k\) defined in Eq.~\eqref{eq: delta_k v_k fully-o pulse}, the excitation density can be estimated using the Landau--Zener formula as
\[
    n_{\mathrm{ex}} = \frac{1}{N} \sum_k \; \exp\left(-\frac{\pi}{2} \frac{\Delta_k^2}{v_k} \right).
\]
This analytical prediction is shown by the dashed lines in Fig.~\ref{fig: densityVsv_linear_anal}, and exhibits excellent agreement with the numerical results across a broad range of drive rates. However, the expression above ceases to be valid in the adiabatic limit (very slow quenches), where one observes a power-law scaling of the excitation density instead of the exponential suppression predicted by the LZ formula. This discrepancy arises because unlike the Landau--Zener protocol, ours is defined over a finite duration. Nonetheless, the approximation remains accurate in the intermediate and fast driving regimes.

In the thermodynamic limit and for intermediate to slow quenches, the excitation probability becomes sharply peaked around \(k = 0\) and \(k = \pi\). This allows us to expand the exponent \(\Delta_k^2/v_k\) near these points and evaluate the excitation density analytically. Expanding around \(k = 0\) (and similarly around \(k = \pi\)), we find up to second order:
\[
    \frac{\Delta_k^2}{v_k} \approx \frac{4\left(f_{o/s}^c + 1\right)^2}{v \left(1 + \dot{f}_{o/s}^{c} \right)} k^2,
\]
Substituting this expansion into the integral for \(n_{\mathrm{ex}}\), we obtain:
\begin{align}
    n_{\mathrm{ex}} & = \frac{2}{\pi} \int_0^{\pi/2} \! dk \;
    \exp\left[-\frac{2\pi \left(f_{o/s}^c + 1\right)^2}{v \left(1 + \dot{f}_{o/s}^{c} \right)} k^2 \right] \nonumber                                           \\
                    & \approx \sqrt{v} \; \frac{\sqrt{1 + \dot{f}_{o/s}^{c}}}{\pi \sqrt{2} \left(f_{o/s}^c + 1\right)}.\label{nex adiabatic noiseless sqrt(v)}
\end{align}

This closed-form expression captures the \(\sqrt{v}\) scaling of the excitation density of finite-size in the intermediate-quench limit and is shown as the dotted lines in Fig.~\ref{fig: densityVsv_linear_anal}.


\section{\label{sec:Noise}Resilience Against Noise}
To study the robustness of the protocol under realistic conditions, we introduce classical noise into the system and examine its effect on the fidelity of state preparation. While in practical implementations, noise with different amplitudes and correlation times across the system, can affect multiple control parameters, we restrict ourselves to two representative forms of uniform noise in the Hamiltonian. Specifically in Eq.~\eqref{eq: Ho} we add: (i) a noisy transverse field of the form \(\xi(t) \sum_i \sigma^z_i\), and (ii) a noisy nearest-neighbor coupling of the form \(\xi(t) \sum_i \sigma^x_i \sigma^x_{i+1}\), where \(\xi(t)\) denotes a stochastic noise process. The latter may be viewed as a perturbation to the implementation of the control function \(f(t)\).

\begin{figure}[t]
    \centering
    \centering
    \includegraphics[width=\linewidth]{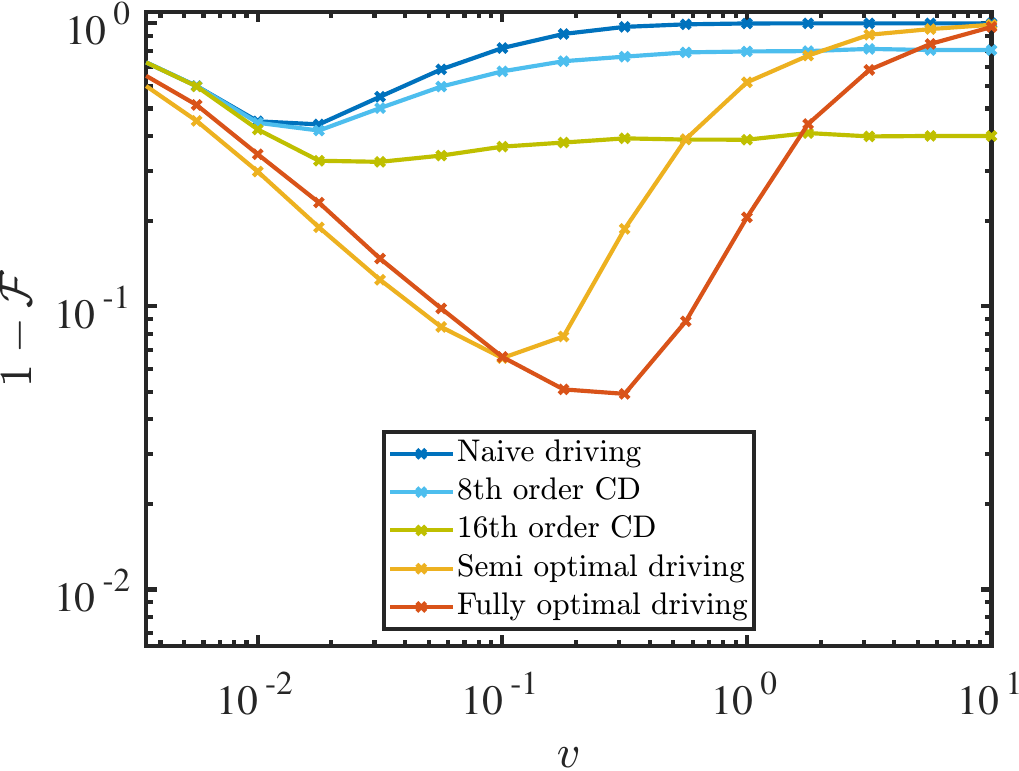}
    \caption{Excitation probability vs.\ drive rate under a linear ramp of the transverse field from $g=10$ (paramagnetic) to $g=0$ (ferromagnetic) in the presence of noise in transverse field, assisted by different protocols. Both semi-optimized and fully optimized protocols retain a significant advantage over the naive protocol and truncated CD.}
    \label{fig: fVsv_Noise}
\end{figure}

The resulting fidelities for the half-quench case are shown in Fig.~\ref{fig: fVsv_Noise}, revealing two notable trends. First, the minimum diabatic error is now achieved at shorter total evolution times (i.e., higher drive rates). Second, despite the presence of noise, the optimized protocol significantly outperforms both the naive ramp and various CD protocols, achieving substantially higher fidelities across a wide range of drive rates.

\subsection{\label{subsec:Analytical_scaling_noise}Analytical Derivation for Density of Excitations in the presence of Noise}

We derive the excitation density in the presence of classical noise for both single-parameter and two-parameter driving, using the optimal pulses obtained in Section~\ref{sec:optimization} for the full-quench case. As in the previous section, we consider weak noise modeled either as a uniform transverse field perturbation, $\xi(t) \sum_i \sigma^z_i$, or as a uniform nearest-neighbor (NN) coupling perturbation, $\xi(t) \sum_i \sigma^x_i \sigma^x_{i+1}$, where $\xi(t)$ denotes stochastic noise with standard deviation \(\sigma_0\) and correlation time \(\tau_c\) described by Ornstein--Uhlenbeck process.

In both cases, the Hamiltonian for each momentum mode acquires a noise-dependent correction, and the  corresponding effective Hamiltonian takes the form
\begin{equation} \label{noise eq: noise Hamiltonian}
    \hat{H}'_k(t) = \frac{v_k t + C_k \xi(t)}{2} \sigma^z + \frac{\Delta_k + S_k \xi(t)}{2} \sigma^x,
\end{equation}
where the coefficients \( C_k \) and \( S_k \) are determined by the nature of the noise. For transverse field noise, which introduces only a longitudinal component, we have \( C_k = 4 \) and \( S_k = 0 \). For the coupling noise, the perturbation includes both longitudinal and transverse components, resulting in \( C_k = 4\cos(k) \) and \( S_k = 4\sin(k) \). Note that the mode-dependent parameters \( v_k \) and \( \Delta_k \) for the naive and optimized protocols are as defined in Sec.~\ref{sec:Analytical_scaling}.
Focusing on the slow-drive regime, noise provides the dominant contribution to the excitation density,
which, for each mode \( k \), is given by
\begin{equation}
    p_k = \frac{\sigma_0^2}{v_k} \left[ C_k^2 R_z(\Delta_k \tau_c) + S_k^2 R_x(\Delta_k \tau_c) \right],
\end{equation}
where the functions \( R_z \) and \( R_x \) are defined in Eq.~\eqref{app: noise formula1}.
A detailed derivation of this expression is provided in Appendix~\ref{LZT with Noise}.

\begin{figure}[t]
    \centering
    \centering
    \includegraphics[width=\linewidth]{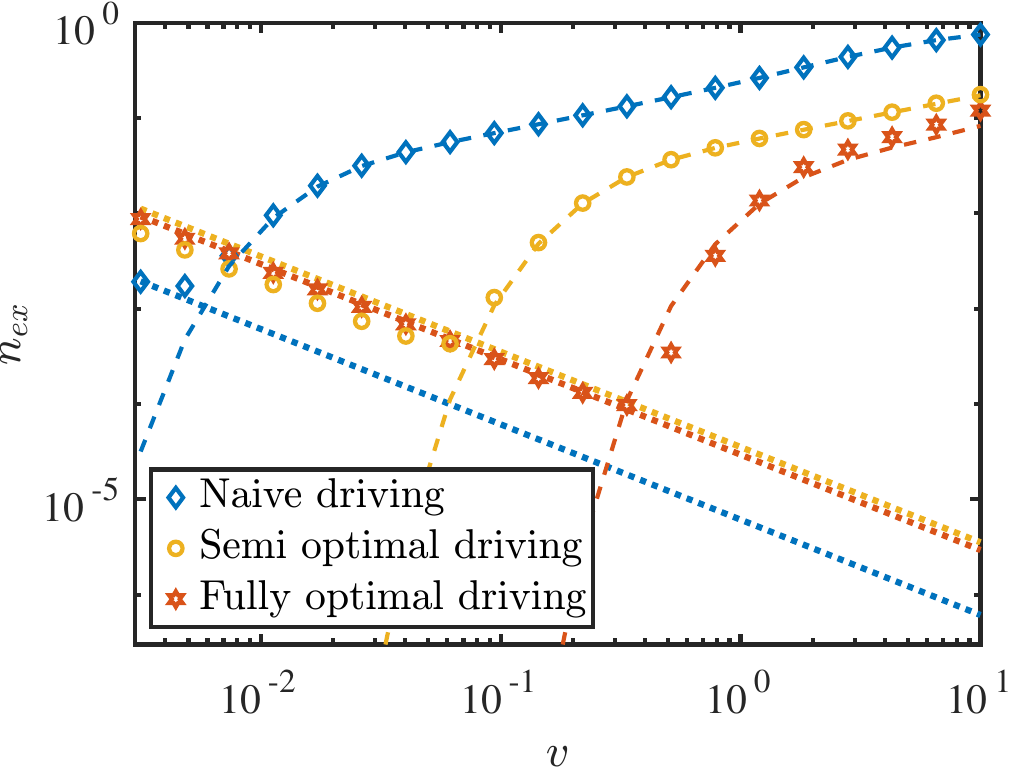}
    \caption{Density of excitations $n_{\mathrm{ex}}$ vs.\ drive rate for different protocols in the fast-noise regime with transverse-field noise. Numerical results are shown as data points. Dashed lines of the same color denote the finite size excitation density noiseless case, while dotted lines indicate the noise-induced contribution. In the fast-drive regime, excitations are primarily generated by drive-induced defects, with numerical results following this trend up to a characteristic drive rate, below which noise effects dominate the contribution. }

    \label{fig: fastnoise_g}
\end{figure}

The resulting density of excitations depends on the structure of the noise term. In the case of a noisy transverse field, the expression reduces to
\begin{equation}\label{app noise: transverse noise}
    \begin{aligned}
        n_{ex}
        \approx & \frac{1}{\pi}\int_0^\pi dk \frac{ (4\sigma_0)^2}{v_k} R_z(\Delta_k \tau_c).
    \end{aligned}
\end{equation}
On the other hand, when the noise affects the nearest-neighbor couplings, both \( R_z \) and \( R_x \) contribute, leading to
\begin{equation} \label{app noise: coupling noise}
    \begin{aligned}
        n_{\mathrm{ex}} = \frac{1}{\pi} \int_0^\pi dk \, \frac{(4\sigma_0)^2}{v_k} \Big[ & \cos^2(k)\, R_z(\Delta_k \tau_c)          \\
                                                                                         & + \sin^2(k)\, R_x(\Delta_k \tau_c) \Big].
    \end{aligned}
\end{equation}
These integrals can be evaluated analytically in the limiting regimes of fast and slow noise.
\begin{figure}[t]
    \centering
    \includegraphics[width=\linewidth]{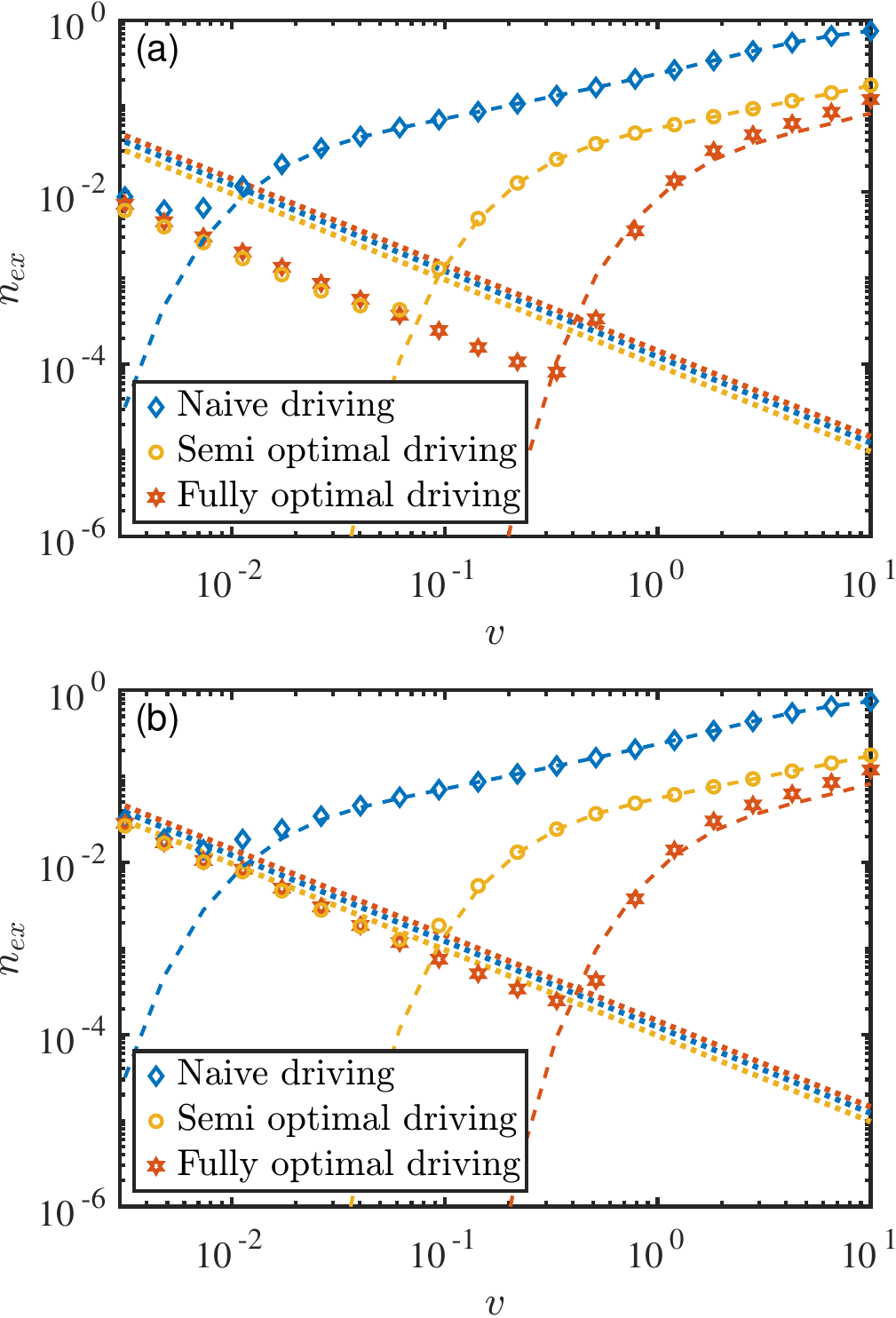}
    \caption{Density of excitations $n_{\mathrm{ex}}$ vs.\ drive rate for different protocols in the fast-noise regime (with noise in the nearest-neighbor coupling). (a) Results for the case $g_i = -g_f = 10$, a noticeable mismatch between numerical data and analytical predictions is observed. (b) Increasing $g_i = -g_f$ from $10$ to $40$ improves agreement between the numerical and analytical results.}
    \label{fig: fastnoise_XX}
\end{figure}
We begin with the fast noise limit, where the correlation time is short compared to the inverse energy scale, i.e., $\tau_c \ll 1/\Delta_k$. In this limit, the noise expressions defined in Eq.~\eqref{app: noise formula1} can be expanded in powers of $\Delta \tau_c$, yielding
\begin{equation}\label{app: noise formula fast noise}
    \begin{aligned}
        R_z(\Delta \tau_c) & \approx \frac{\pi}{2} \Delta \tau_c + \mathcal{O}(\Delta^2\tau_c^2),                  \\
        R_x(\Delta \tau_c) & \approx \frac{\pi}{2} \left(1 - \Delta \tau_c\right) + \mathcal{O}(\Delta^2\tau_c^2).
    \end{aligned}
\end{equation}

We ignore the higher-order terms and evaluate the density of excitations in the presence of transverse field noise under the naive driving protocol. Substituting $v_k = 4v$ and $\Delta_k = 4\sin(k)$ into Eq.~\eqref{app noise: transverse noise}, we obtain
\begin{equation}
    n_{\mathrm{ex}} \approx \frac{16\sigma_0^2\tau}{v}.
\end{equation}
For the case of optimized driving in the presence of transverse field noise, the excitation density can be calculated by substituting the expressions for $v_k$ and $\Delta_k$ from Eq.~\eqref{eq: delta_k v_k fully-o pulse} into the integral in Eq.~\eqref{app noise: transverse noise}. Using the fast-noise limit approximation for $R_z(\Delta \tau_c)$ from Eq.~\eqref{app: noise formula fast noise}, and performing the integration, we obtain
\begin{equation}
    n_{\mathrm{ex}} \approx \frac{4\sigma_0^2 \tau_c}{v} \, \beta_{o/s} \left(1 - \frac{\mathcal{S}^{-1}_{o/s}(\sqrt{|\gamma_{o/s}|})}{\sqrt{|\gamma_{o/s}|}} + \frac{\gamma_{o/s}}{3} \right),
\end{equation}
where the function \( \mathcal{S}^{-1}_o \) denotes \( \sin^{-1} \), corresponding to the fully optimal case with \( -1 < \gamma_o < 0 \), while \( \mathcal{S}^{-1}_s \) denotes \( \sinh^{-1} \), applicable in the semi-optimal case where \( \gamma_s > 0 \).

We next consider the case of noise in the nearest-neighbor coupling. For naive driving, substituting Eq.~\eqref{app: noise formula fast noise} into Eq.~\eqref{app noise: coupling noise}, we obtain
\begin{equation}
    n_{ex} \approx \frac{\sigma_0^2}{v} \left( \pi - \frac{16\tau_c}{3} \right).
\end{equation}

For optimized protocols, substituting corresponding expression of $v_k$ and $\Delta_k$ expressions into Eq.~\eqref{app: noise formula fast noise} and then into Eq.~\eqref{app noise: coupling noise}, we obtain
\begin{equation}
\begin{aligned}
    n_{ex} \approx &\frac{4\sigma_0^2}{v} \bigg[  \tau_c\, \beta_{o/s} \left(1 - \frac{\mathcal{S}^{-1}_{o/s}\!\left(\sqrt{|\gamma_{o/s}|}\right)}{\sqrt{|\gamma_{o/s}|}} - \frac{|\gamma_{o/s}|}{3} \right) \\
    & + \left( K\!\left( \frac{|\gamma_{o/s}|}{|\gamma_{o/s}|+1} \right) - E\!\left( \frac{|\gamma_{o/s}|}{|\gamma_{o/s}|+1} \right) \right) \frac{\sqrt{|\gamma_{o/s}|+1}}{|\gamma_{o/s}|} \bigg],
\end{aligned}
\end{equation}
where $K$ and $E$ denote the complete elliptic integrals of the first and the second kind, respectively.

Figures~\ref{fig: fastnoise_g} and~\ref{fig: fastnoise_XX} demonstrate that the optimized protocol suppresses excitations by nearly two orders of magnitude compared to the naive protocol at its optimal drive rate. They also show that, for noise of equal strength and correlation time, its impact is more significant when acting on the couplings than on the transverse field. As we will discuss later this trend does not hold in the regime of long correlation times.

The analytical expressions agree well with the numerical data, particularly for noise in the transverse field.
In the case of noise in the nearest-neighbor coupling, some deviation arises due to the fact that the analytical derivations assume an infinite drive window, while the numerical simulations are performed over a finite range. As shown in Fig.~\ref{fig: fastnoise_XX}b, the agreement improves as the drive range is extended, consistent with the assumptions underlying the analytical treatment.

We now turn to the slow noise limit, where the correlation time $\tau_c$ is much larger than the
system size $N$ so that $\tau_c \gg 1/\Delta_k$ for all $k$ modes (as $\Delta_k > \Delta_{k_0} \sim 1/N$). In this regime, the noise varies slowly compared to the dynamics of the system, and therefore the noise terms in Eq.~\eqref{app: noise formula1} can be expanded in powers of $1/(\Delta \tau_c)$. This gives
\begin{equation}\label{app: noise formula slow noise}
    \begin{aligned}
        R_z(\Delta \tau_c) \approx R_x(\Delta \tau_c) \approx \frac{\pi}{4 \Delta \tau_c} + \mathcal{O}\left(\frac{1}{\Delta^2 \tau_c^2}\right).
    \end{aligned}
\end{equation}
in what follows we neglect the higher-order terms $\mathcal{O}(1/\Delta^2 \tau_c^2)$. Since $R_z$ and $R_x$ become equal in this limit, both transverse field noise and coupling noise lead to the same expression for the density of excitations. Substituting into Eq.~\eqref{app noise: transverse noise} or Eq.~\eqref{app noise: coupling noise}, we obtain
\begin{equation}
    n_{ex} \approx \frac{4\sigma_0^2}{\tau_c} \int_{k_0}^{\pi - k_0} dk \, \frac{1}{\Delta_k v_k}. \label{app noise: slow noise}
\end{equation}
We again remind that the above and the following expressions are valid only when the system size is much smaller than the correlation time (but large enough for the density of excitations to be approximated as an integral).

In the case of naive driving, the above integral can be evaluated exactly, giving
\begin{equation}
    n_{ex} \approx \frac{\sigma_0^2}{v \tau_c} \frac{\log(2/k_0)}{2}.
\end{equation}

For the diabatic protocol, the expression for $\Delta_k$ is more involved. To simplify the integral, we expand the coefficient of $\sin(k)$ in $\Delta_k$, given by $\beta \sec^2(k) \left( \sqrt{1 + \gamma \cos^2(k)} - 1 \right)$, around $k = 0$. To leading order, this becomes the constant $4(f_{o/s}^c + 1)$. This approximation allows the integral in Eq.~\eqref{app noise: slow noise} to be evaluated as
\begin{equation}
    n_{ex} \approx \frac{\sigma_0^2}{v \tau_c} \cdot \frac{\tanh^{-1} \left( \frac{\sqrt{1 + \gamma_{o/s}} \cos(k_0)}{\sqrt{1 + \gamma_{o/s} \cos^2(k_0)}} \right)}{2(f_{o/s}^c + 1)\sqrt{1 + \gamma_{o/s}}}.
\end{equation}

\begin{figure}[t]
    \centering
    \includegraphics[width=\linewidth]{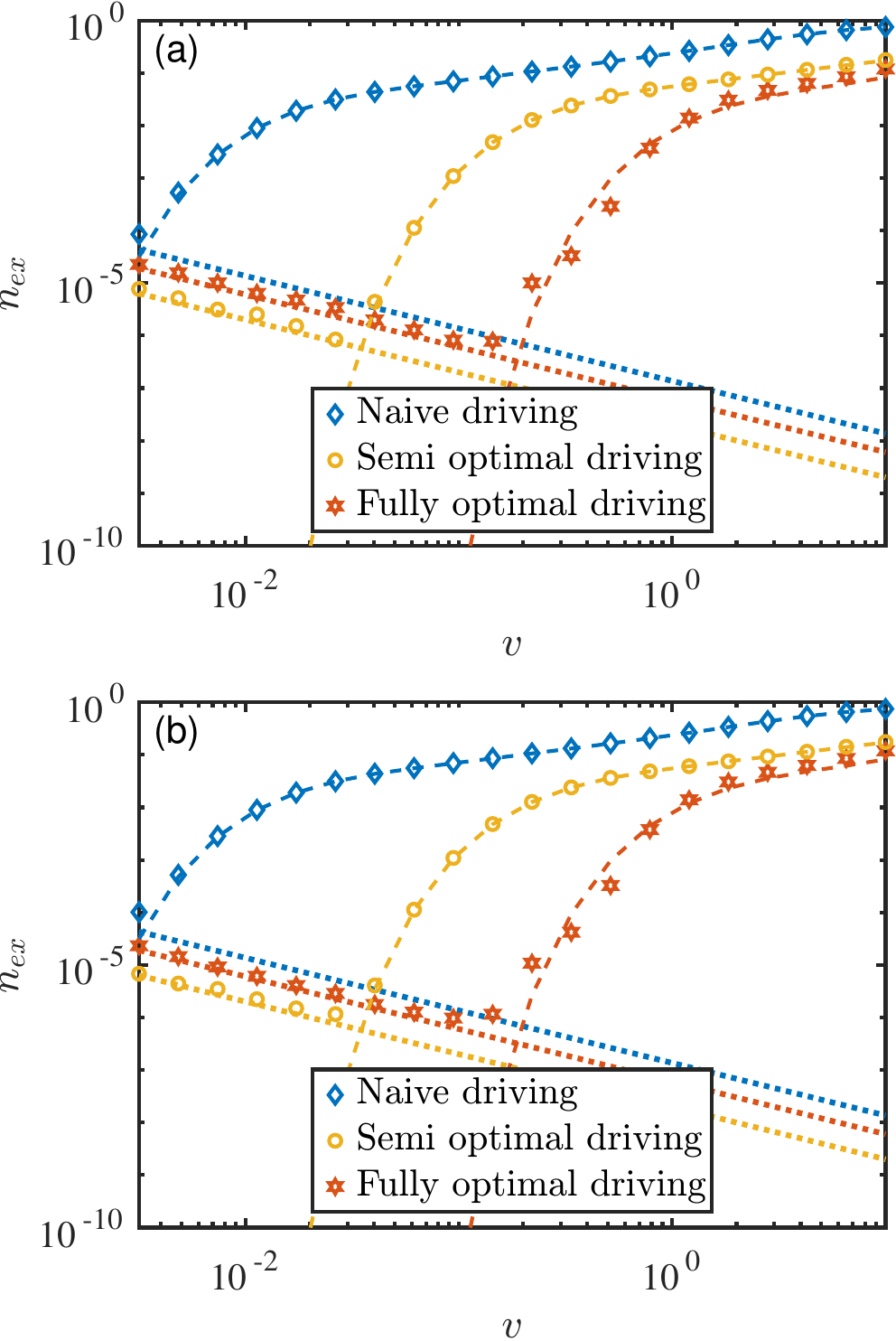}
    \caption{Density of excitations $n_{\mathrm{ex}}$ under noise with a long correlation time ($500$ units). Panel (a) corresponds to transverse-field noise, while panel (b) corresponds to the noise in nearest neighbor coupling.}
    \label{fig:slownoise}
\end{figure}

\vspace{1\baselineskip}
As shown in Fig.~\ref{fig:slownoise}, the analytical results agree well with numerical simulations. Moreover, the excitation density in this regime is insensitive to whether the noise acts on the transverse field or on the coupling terms.

\section{\label{sec:conclusion}Conclusion}
\vspace{1\baselineskip}

In this work, we employed the QB framework to derive an optimal drive protocol that suppresses excitations in a quantum critical system using only local auxiliary terms in the Hamiltonian. In scenarios where CD approaches—particularly truncated CD—fail to achieve the desired fidelity or become impractical due to experimental considerations, QB provides a viable alternative. Moreover, it preserves a key advantage of the CD formalism: a standardized procedure for deriving optimal protocols applicable to any quantum critical system.

To demonstrate the effectiveness of this approach, we considered the transverse field Ising model driven linearly across the quantum critical point. While adiabatic evolution can, in principle, yield high-fidelity final states, it requires impractically slow driving rates, rendering naive driving ineffective. This limitation persists even after incorporating several orders of CD corrections, which fail to achieve high fidelity in both the slow and fast driving regimes.
To overcome this, we modified the Hamiltonian by introducing a second control parameter $f(t)$ coupled to the longitudinal terms. Analysis of the critical modes near \( k_0 \approx 0 \) allowed us to identify a simple quadratic form for $f(t)$ which improves the fidelity for all drive rates.
We next employed a systematic approach to determine a fully optimal trajectory in the control space by minimizing the temporal variance of the corresponding CD Hamiltonian (of the total modified Hamiltonian). This later strategy is inspired by the QB formulation.

Using this protocol, we find that the excitation probability retains its exponential dependence on the drive rate, but with a significantly improved decay exponent compared to naive driving. To highlight its advantage, we considered a case with weak noise of strength $\sim 1/N$ and unit correlation time, where naive and approximate CD protocols yield a maximum fidelity of about $0.6$, whereas our optimized protocol achieves fidelity values as high as $0.96$ under the same conditions.

The optimized $f(t)$ exhibits a near-quadratic form in the vicinity of the critical time, this structure allows us to derive analytical expressions for the effective gap and drive rate for each mode and approximate the excitation probability using the Landau-Zener model. The resulting predictions for excitation density match numerical simulations well.
We extended the analysis to the case where noise is present in both the transverse field and the longitudinal coupling.
In both the naive and optimized protocols, we find that for weak noise, the excitation density scales as $1/v$, with coefficients dependent on the noise strength, correlation time, and the effective drive parameters. In the slow noise limit ($\tau_c \gg N$), we find that transverse field and coupling noise have equivalent effects. However, in the fast noise regime ($\tau_c \ll 1$), coupling noise leads to stronger degradation in fidelity than transverse field noise.
Combined with the density of excitations for a noiseless drive, one can readily estimate the optimal drive rate in the presence of weak noise in the transverse field and coupling term.
Although our analysis centers on the TFIM, the methods presented here are readily applicable to a wider class of free-fermion systems.

\begin{acknowledgments}
	We thank Professor Daniel Loss for many illuminating discussions and for the hospitality extended to S.G. during the visit to Basel. D.S. would like to thank CSIR for funding by their file
	no. 09/1020(0216)/2021-EMR-I.
\end{acknowledgments}

\newpage
\onecolumngrid
\appendix
\section{Conterdiabtic formalism for finite size chain}
\label{app:CD}

Here, we review the results from \cite{delcampo2014IsingModel,damski2014counterdiabatic} for a TFIM described by the Hamiltonian in  Eq.~\eqref{Models eq: model TFIM}, which decomposes into a set of independent two-level systems, one for each
$k$-mode. The CD term can thus be computed separately for each mode and subsequently expressed in the full many-body basis by sandwiching it between the corresponding Nambu spinors,
\begin{equation}\label{eq: CD XY Hamiltonian}
    H_{CD} = -\dot{g}~ \sum_{k>0} f_k(g)~\hat{\psi}_k^\dagger~ \tau^y ~\hat{\psi}_k,
\end{equation}
where $f_k(g) = \frac{1}{2} \sin{k} [(g - J \cos k)^2 + (J \sin k)^2]^{-1}$. An Fourier transform to real-space fermionic operators, followed by an Jordan-Wigner transformation yields
\begin{equation}\label{eq: CD real space Hamiltonian app}
    \hat{H}_{CD}/\dot{g} = \sum_{m=1}^{N/2 - 1} h_m(g)~\mathcal{H}^{[m]} + \frac{1}{2} h_{N/2}(g)~\mathcal{H}^{[N/2]},
\end{equation}
where $h_m(g) = \frac{1}{N} \sum_{k>0} f_k(g) \sin(mk)$ and
\[
    \mathcal{H}^{[m]} = \sum_{n=1}^{N} \left[\sigma_n^x \sigma_{n+m}^y + \sigma_n^y \sigma_{n+m}^x\right] \left(\prod_{j=n+1}^{n+m-1} \sigma_j^z\right).
\]
This CD Hamiltonian involves long-range interactions extending up to $N/2$ sites (i.e., 25 sites for $N=50$), which poses a challenge for experimental implementation — a common limitation of counterdiabatic protocols in many-body systems.

To address this, one can consider a truncated CD Hamiltonian including only short-range interactions:
\begin{equation}
    h_m'(g) =
    \begin{cases}
        h_m(g), & \text{if } 1 \le m \le M \\
        0,      & \text{if } m > M
    \end{cases}
\end{equation}
where $M$ is the maximum interaction range. The truncated Hamiltonian is then obtained by replacing $h_m(g)$ with $h_m'(g)$ in Eq.~\eqref{eq: CD real space Hamiltonian app}.

Using the exact form of $h_m(g)$ for a finite system size $N$, as computed in \cite{damski2014counterdiabatic}, the performance of the approximate CD protocol truncated at various $M$ can be numerically evaluated.

\section{Landau Zener Drive with a Noisy field}\label{LZT with Noise}
We follow the approach of Ref.~\cite{krzywda2020adiabatic} wherein longitudinal noise was considered. We extend the analysis to include
transverse noise as well. The full two level Hamiltonian with noise terms included is given by:
\begin{equation}\label{app noise eq: noise Hamiltonian}
    \Hhat(t) = \frac{\epsilon (t) + C ~\xi (t)}{2} \sigmaz + \frac{\Delta+S ~\xi(t)}{2} \sigmax = \frac{1}{2} \mbf{r}(t)\cdot \vec{\sigma},
\end{equation}
where $C$ and $S$ are model dependent constants.
We define $\U(t) =~ \ket{g(t)}\bra{-}~ +~  \ket{e(t)}\bra{+}$ where $\{\ket{g(t)}, \ket{e(t)}\}$ and $\{\ket{-}, \ket{+}\}$ are the ordered eigenbasis of $\Hhat(t)$ and $\sigmaz$, respectively.  Plugging $\ket{\psi(t)}=\U(t) \ket{\psi'(t)}$ into the Schrodinger equation for $\Hhat$ we obtain:
\begin{equation}\label{noisy eq: adiabatic SC eqn}
    i \h \partial_t \kp{'(t)} = \left( \U_t^{\+} \Hhat(t) \U_t - i \U_t^\+ \partial_t \U_t \right) \kp{'(t)},
\end{equation}
where $\U_t^\+ \Hhat(t) \U_t = \frac{1}{2} {r}(t) \sigmaz$.  From the definition of the instantaneous eigenstates, the term $ i \U_t^\+ \dot{\U}_t $  can be evaluated as $-\dot{\theta}/2~ \sigmay$, where $\hat{r}=[\sin\theta(t),0,-\cos\theta(t)]$ and $\theta(t)=-\cot^{-1} (\frac{\epsilon (t) + C ~\xi (t)}{\Delta+S ~\xi(t)})$ is the angle subtended by $\mbf{r}(t)$ from the negative Z axis.
For small noise, we can perform the following approximations:

\begin{equation}
    \begin{aligned}
        r(t) =                     & \sqrt{(\epsilon (t) + C ~\xi (t))^2+(\Delta+S ~\xi(t))^2}                                                                                                      \\
        \approx                    & \underbrace{\sqrt{\epsilon (t)^2+\Delta^2}}_{r_0(t)} + \underbrace{\frac{C~\epsilon(t) + S~\Delta}{r_0(t)}}_{-A(t)} \xi(t)                            \\
        \cot~\theta(t)=           & -\frac{\epsilon (t) + C ~\xi (t)}{\Delta+S ~\xi(t)}                                                                                                            \\
        \approx                    & -\frac{\epsilon(t)}{\Delta} + \frac{\xi(t)}{\Delta}\left(-C+\frac{S~\epsilon(t)}{\Delta}\right)                                                                \\
        \implies \theta(t) \approx & ~\underbrace{\arccot \frac{-\epsilon (t)}{\Delta}}_{\vartheta(t)}+ \underbrace{\frac{C \Delta - S \epsilon(t)}{r_0(t)}}_{B(t)} \frac{\xi(t)}{r_0(t)},
    \end{aligned}
\end{equation}
where the approximation $\arccot(x+\delta)\approx \arccot(x)-\frac{\delta}{1+x^2}$ (for small $\delta$) has been used in the last line. In the above, $r_0(t)$, $A(t)$, $B(t)$ and $\vartheta(t)$ have been defined for convenience. 
Thus the Hamiltonian in the adiabatic basis (Eq. \ref{noisy eq: adiabatic SC eqn}) is approximated as:
\begin{equation}\label{noisy eq: adiabatic Hamil}
    \begin{aligned}
        \Hhat^{ad}(t) = & \frac{r(t)}{2}  \sigmaz + \frac{\dot{\theta}}{2}~ \sigmay                                       \\
        \approx         & \frac{r_0(t)-\xi_\parallel (t)}{2} \sigmaz +  \frac{\dot{\vartheta}_t+\xi_\perp (t)}{2} \sigmay,
    \end{aligned}
\end{equation}
where
\begin{equation}
    \begin{aligned}
        \xi_\parallel (t) = & A(t)~ \xi(t)                                           \\
        \xi_\perp (t) =     & \frac{d}{dt} \left[B(t)~ \frac{\xi(t)}{r_0(t)}\right]
    \end{aligned}
\end{equation}
Next, we shall write equation of motion of the form $\kp{'(t)}=c_+(t) e^{-i\phi(t)/2} \ket{+} + c_-(t) e^{i\phi(t)/2} \ket{-}$, where $\phi(t)=\int_{-T}^{t} r(t')~dt'$, and evolve this with the adiabatic Hamiltonian in Eq \ref{noisy eq: adiabatic Hamil}.  We now consider the system initialized in the lowest energy state (i.e., $c_-\left(-\infty\right)=1,~ c_+\left(-\infty\right)=0$). We look for the solution in a perturbative way, by writing $c_\pm(\tau ) = c^0_\pm (\tau ) + \lambda c^1_\pm(\tau ) + \lambda^2 c^2_\pm(\tau ) + ...$ where $\lambda$ counts the powers of noise, i.e., we replace $\xi_{\perp/\parallel}$ by $\lambda\xi_{\perp/\parallel}$  and substitute in the equation of motion for $c_+(t):$
\begin{equation}
    \dot{c}_+(t)= i  \frac{\xi_\parallel(t) c_+(t)}{2}  -   \frac{\dot{\vartheta}_t+\xi_\perp (t)}{2} e^{i \int_{-T}^{t} r(t')~dt'} c_-(t).
\end{equation}
In doing so we obtain the following first order equation:
\begin{equation}
    c_+^1(T)=-\frac{1}{2}\int_{-T}^{T}  dt~  \xi_\perp (t) e^{i \int_{-T}^{t} r(t')~dt'} c_-(t)
\end{equation}

\begin{equation}
    \begin{aligned}
        |c_+^{(1)}(\infty)|^2 = &
        \frac{1}{4} \int_{-\infty}^{\infty} d\tau_1 d\tau_2 \langle \xi_{\perp}(\tau_1) \xi_{\perp}(\tau_2) \rangle e^{i \int_{\tau_2}^{\tau_1} r_0(\tau) d\tau}                            \\
        \approx                 \frac{1}{4} \int_{-\infty}^{\infty} &d\tau_1 d\tau_2~  B(\tau_1)~B(\tau_2)~\frac{\langle \dot{\xi}(\tau_1) \dot{\xi}(\tau_2) \rangle}{
            r_0(\tau_1) r_0(\tau_2)} e^{i \int_{\tau_2}^{\tau_1} r_0(\tau) d\tau}.
        \label{noisy eq: B1}
    \end{aligned}
\end{equation}
In the above, we retained only the term proportional to the noise derivative, 
$\xi_{\perp}(\tau) \sim B(\tau)\, \dot{\xi}/r$, and neglected the static contribution 
$\propto \xi(\tau)$, since slow noise is not expected to generate excitations.  
 Now for an Ornstein-Uhlenbeck process,
\begin{equation}
    \langle \dot{\xi}(\tau_1) \dot{\xi}(\tau_2) \rangle = \frac{\sigma_0^2}{\tau_c} \left( 2 \delta(\tau_1 - \tau_2) - \frac{1}{\tau_c} e^{-\frac{|\tau_1 - \tau_2|}{\tau_c}} \right),
\end{equation}
as derived in Ref~\cite{krzywda2020adiabatic}. In Fourier space, this can be written as
\begin{equation}
    2\delta(\tau_1 - \tau_2) - \frac{1}{\tau_c} e^{-\frac{|\tau_1 - \tau_2|}{\tau_c}} = \int_{-\infty}^{\infty} \frac{d\omega}{2\pi} \frac{2\omega^2 \tau_c^2}{1 + \omega^2 \tau_c^2} e^{i\omega(\tau_1 - \tau_2)},
\end{equation}
and then substituted into Eq. \ref{noisy eq: B1}:
\begin{equation}
\begin{aligned}
    \frac{\sigma_0^2}{4\tau_c} \int_{-\infty}^{\infty} &\frac{d\omega}{2\pi} \frac{2\omega^2 \tau_c^2}{1 + \omega^2 \tau_c^2} \left[ \int_{-\infty}^{\infty} d\tau_1 \frac{B(\tau_1)}{r_0(\tau_1)} e^{i\omega \tau_1 + i \int_{0}^{\tau_1} r_0(\tau) d\tau} \right]\\
    &\left[ \int_{-\infty}^{\infty} d\tau_2 \frac{B(\tau_2)}{r_0(\tau_2)} e^{-i\omega \tau_2 - i \int_{0}^{\tau_2} r_0(\tau) d\tau} \right].
\end{aligned}  
\end{equation}
Using the stationary phase method,
\begin{equation}
    \int g(\tau) e^{ih(\tau)} d\tau \approx \sum_{\tilde{\tau}  \text{~s.t.~} h'(  \tilde{\tau})=0 } g(\tilde{\tau}) e^{i h(\tilde{\tau})} \int e^{i h''(\tilde{\tau}) \frac{(x - \tilde{\tau})^2}{ 2}} dx,
\end{equation}
where \( \tilde{\tau} \) is found from the equation
\begin{equation}
    \partial_{\tau} (\omega \tau +\int_{0}^{\tau} d\tau' r_0(\tau') ) = 0 \Rightarrow \omega =- r_0(\tilde{\tau}),
\end{equation}
and reads \( \tilde{\tau} = \pm \frac{\sqrt{\omega^2 - \Delta^2}}{v} \). The second derivative $\partial_{\tau}^2 (\omega \tau +\int_{0}^{\tau} d\tau' r_0(\tau') )$ is
\begin{equation}
    h''(\pm \tilde{\tau}) = \pm \frac{v}{\omega} \sqrt{\omega^2 - \Delta^2}.
\end{equation}
As a result,
\begin{equation}
    \begin{aligned}
        &\int_{-\infty}^{\infty}  d\tau_1 B(\tau_1) \frac{e^{i\omega \tau_1 + i \varphi(\tau_1)}}{r_0(\tau_1)}  \\
        \approx  &B(\tilde{\tau})\frac{e^{i\omega \tilde{\tau} +i\varphi(\tilde{\tau})}}{r_0(\tilde{\tau})}
        \int_{-\infty}^{\infty} d\tau_1 \, e^{i h''(\tilde{\tau}) \frac{(\tau_1 - \tilde{\tau})^2}{2}}
        + (\tilde{\tau} \to -\tilde{\tau})                                                                                                                                                                                                                                                                                                   \\
        =                       & \sqrt{\frac{2 \omega \pi}{v \sqrt{\omega^2-\Delta^2}}} \left[ (C \Delta - S v\tilde{\tau}) \frac{e^{i\omega \tilde{\tau} +i\varphi(\tilde{\tau})-i\pi/4}}{r_0^2(\tilde{\tau})} + (C \Delta + S v\tilde{\tau}) \frac{e^{-i\omega \tilde{\tau} -i\varphi(\tilde{\tau})+i\pi/4}}{r_0^2(\tilde{\tau})} \right] \\
        =                       & \sqrt{\frac{8 \pi}{v \omega^3\sqrt{\omega^2-\Delta^2}}} \left[C \Delta \cos \left( \omega \tilde{\tau} + \varphi(\tilde{\tau}) - \frac{\pi}{4} \right) -i ~S v \tilde{\tau} ~ \sin \left( \omega \tilde{\tau} + \varphi(\tilde{\tau}) - \frac{\pi}{4} \right) \right],
    \end{aligned}
\end{equation}

where \( \varphi(\tilde{\tau}) = \int_{0}^{\tilde{\tau}} r_0(\tau') d\tau' \). The second integral is the complex conjugate, leading to
\begin{equation}
    \begin{aligned}
        2\pi \frac{\sigma_0^2 }{v\tau_c} \int_{\Delta}^{\infty} \frac{d\omega}{2\pi} \frac{2\omega^2 \tau_c^2}{1 + \omega^2 \tau_c^2} \frac{1}{\omega^4 \sqrt{ 1 - \frac{\Delta^2}{\omega^2}}} & \bigg(C^2  \Delta^2 \cos^2  \left( \omega \tilde{\tau} + \varphi(\tilde{\tau}) - \frac{\pi}{4} \right)              \\
                                                                                                                                                                                               & + S^2 v^2 \tilde{\tau}^2 ~ \sin^2 \left( \omega \tilde{\tau} + \varphi(\tilde{\tau}) - \frac{\pi}{4} \right) \bigg).
    \end{aligned}
\end{equation}
Neglecting the fast oscillatory terms,
\begin{equation}
    \begin{aligned}
        |c_+^{(1)}(\infty)|^2 \approx & \frac{\sigma_0^2 }{v\tau_c} \int_{\Delta}^{\infty} d\omega \frac{\omega^2 \tau_c^2}{1 + \omega^2 \tau_c^2} \frac{1}{\omega^4 \sqrt{ 1 - \frac{\Delta^2}{\omega^2}}}  \left(C^2 \Delta^2  + ~S^2 (\omega^2 - \Delta^2)  \right) \\
        (\omega\rightarrow x/\tau_C)= & \frac{\sigma_0^2 \tau_c ^2}{v} \int_{\Delta \tau_c}^\infty dx \frac{1}{x^2+x^4} \frac{1}{\sqrt{1-(\Delta \tau_c/x)^2}}  \left(C^2 \Delta^2  + ~(S/\tau_c)^2 (x^2 - (\Delta \tau_c)^2)  \right)                                \\
        =                             & \frac{\sigma_0^2}{v} \int_{\Delta \tau_c}^\infty dx \frac{1}{x^2+x^4}   \left( \frac{C^2 (\Delta \tau_c)^2}{\sqrt{1-(\Delta \tau_c/x)^2}}  + ~S^2 x^2 \sqrt{1-(\Delta \tau_c/x)^2 } \right).
    \end{aligned}
\end{equation}
Defining
\begin{equation}\label{app: noise formula1}
    \begin{aligned}
        R_z(\Delta \tau_c) = & (\Delta \tau_c)^2 \int_{\Delta \tau_c}^{\infty} dx \frac{(1 - \frac{\Delta^2 \tau_c^2}{x^2})^{-1/2}}{x^2 + x^4} = \frac{\pi}{2} \Delta \tau_c \left( 1 - \frac{1}{\sqrt{1 + 1/(\Delta^2 \tau_c^2)}} \right) \\
        R_x(\Delta \tau_c) = & \int_{\Delta \tau_c}^\infty dx \frac{x^2 (1 - \frac{\Delta^2 \tau_c^2}{x^2})^{1/2}}{x^2+x^4} =  \frac{\pi}{2}  \sqrt{(\Delta \tau_c)^2+1} \left( 1 - \frac{1}{\sqrt{1 + 1/(\Delta^2 \tau_c^2)}} \right),
    \end{aligned}
\end{equation}
we obtain the final result:
\begin{equation}\label{app: noise formula2}
    |c_+^{(1)}(\infty)|^2 = \frac{ \sigma_0^2}{v} (C^2 R_z(\Delta \tau_c)+ S^2 R_x(\Delta \tau_c)).
\end{equation}

\section{Variance of Counterdiabatic Hamiltonian} \label{app:Hilbert}
As shown in Appendix~\ref{app:CD}, the CD Hamiltonian for the Ising model decomposes into a set of decoupled two-level Hamiltonians, one for each $k$-mode,  
$
\hat{H}_k = -\dot{g}\, f_k(g)\, \hat{\sigma}^y_k .
$
Since the ground state is a product of the ground states of individual $k$-modes, and the expectation value of $\hat{H}_k$ with respect to the ground state vanishes, we have
$
\bra{\psi_g}\hat{H}'_k \hat{H}'_{k'}\ket{\psi_g} 
= \langle \hat{H}'_k \rangle \langle \hat{H}'_{k'} \rangle = 0 .
$
Moreover,
\[
\hat{H}_k^2 = \big(-\dot{g}\, f_k(g)\, \hat{\sigma}^y_k\big)^2 
= \dot{g}^2 f_k^2(g) \, \mathbb{I} ,
\]
so that the variance of the full CD Hamiltonian with respect to the instantaneous ground state is
\begin{equation}
\sigma^2(\hat{H}_{\mathrm{CD}}, \ket{\psi_g})
= \langle \hat{H}_{\mathrm{CD}}^2 \rangle - \langle \hat{H}_{\mathrm{CD}} \rangle^2
= \dot{g}^2 \sum_k f_k^2(g).
\end{equation}
Similarly, the square of Hilbert--Schmidt norm of $\hat{H}_{\mathrm{CD}}$ follows from
$$ ||\hat{H}_{CD}||^2 = \mathrm{Tr}(\hat{H}_{\mathrm{CD}}^2)
= \sum_k \mathrm{Tr}(\hat{H}_k'^2)
+ \sum_{k\neq k'} \mathrm{Tr}(\hat{H}'_k \hat{H}'_{k'}).$$
Since
$\mathrm{Tr}(\hat{H}_k'^2) = 2 \dot{g}^2 f_k^2(g)$, and
$\mathrm{Tr}(\hat{H}'_k \hat{H}'_{k'}) 
= \mathrm{Tr}(\hat{H}'_k)\, \mathrm{Tr}(\hat{H}'_{k'}) = 0$, we obtain
\begin{equation}
||\hat{H}_{CD}||^2=\mathrm{Tr}(\hat{H}_{\mathrm{CD}}^2)
= 2 \dot{g}^2 \sum_k f_k^2(g).
\end{equation}

\twocolumngrid
\bibliographystyle{apsrev4-2}
\bibliography{QB.bib}

\end{document}